\documentclass{PoS}
\usepackage{amsmath}
\usepackage{amssymb} 

\newcommand{\be}{\begin{equation}}
\newcommand{\ee}{\end{equation}}
\newcommand{\bea}{\begin{eqnarray}}
\newcommand{\eea}{\end{eqnarray}} 
\newcommand{\bel}{\begin{align}}
\newcommand{\eel}{\end{align}}

\title{Operators for scattering of particles with spin }

\ShortTitle{Operators for scattering of particles with spin }

\author{\speaker{Sasa Prelovsek}%
       \\
       Department of Physics, University of Ljubljana, 1000 Ljubljana, Slovenia\\
       Jozef Stefan Institute, 1000 Ljubljana, Slovenia\\
       Theory Center, Jefferson Lab, 12000 Jefferson Avenue, Newport News, Virginia 23606, USA \\
      E-mail: \email{sasa.prelovsek@ijs.si}}

\author{Ursa Skerbis\\
         Jozef Stefan Institute, 1000 Ljubljana, Slovenia\\\
        E-mail: \email{ursa.skerbis@ijs.si}}
        
        \author{C.B. Lang\\
         Institute of Physics,  University of Graz, A--8010 Graz, Austria \\\
        E-mail: \email{christian.lang@uni-graz.at}}

\abstract{Operators for simulating the scattering of two particles with spin are constructed. Three methods are shown to give the consistent lattice operators for $PN$, $PV$, $VN$ and $NN$ scattering, where $P$, $V$ and $N$ denote pseudoscalar meson, vector meson and nucleon. The projection method   leads to one or several operators $O_{\Gamma,r,n}$ that transform according to a given irreducible representation $\Gamma$ and row $r$.   However, it gives little guidance on which continuum quantum numbers of total  $J$, spin $S$, orbital  momentum $L$ or single-particle helicities $\lambda_{1,2}$ will be related with a given operator. This is remedied with the helicity and partial-wave   methods. There first the operators with good continuum quantum numbers  $(J,P,\lambda_{1,2})$ or $(J,L,S)$ are constructed and then subduced to the irreps $\Gamma$ of the discrete lattice group.  The results indicate which linear combinations $O_{\Gamma,r,n}$ of various $n$ have to be employed in the simulations in order to enhance couplings to the states with desired continuum quantum numbers.  The total momentum of two hadrons is restricted to zero since parity $P$ is a good   quantum number in this case.   
   }

\FullConference{34th annual "International Symposium on Lattice Field Theory"\\
       24-30 July 2016\\
           University of Southampton, UK}

\begin{document}

\section{Introduction}

Most of hadrons, particularly the exotic ones, are resonances that  appear as cross-section peaks in the strong scattering of lighter hadrons. This requires  the study of two-hadron interactions by simulating two-hadron scattering on the lattice.  The first  step is to build the  operators that create and annihilate the two-hadron system of the desired quantum numbers. Channels with  two spinless hadrons  have been extensively studied. The simulations of two-hadron systems where one or both hadrons carry non-zero spin focused mostly on partial wave $L=0$.   There is a great need for lattice results on further channels or higher partial waves  in this case. 

We construct  the  lattice operators for two-hadron scattering, where one or both hadrons carry spin and show that 
three independent methods lead to consistent results. The more detailed presentation \cite{Prelovsek:2016iyo} provides also the  proofs of the methods,  explicit expressions for operators and  all necessary details to construct them.    
We consider channels that involve the  nucleon $N$ or/and vectors $V=J/\psi,~\Upsilon_b,~D^*,B^*,...$ which are (almost) stable under strong interactions. The $PN$ or $VN$ scattering is crucial for ab-initio study of baryon resonances and pentaquark candidates,  $PV$ is essential for mesonic resonances and tetraquark candidates, while $NN$
is needed  to grasp two-nucleon interaction and deuterium.  The   periodic  boundary conditions in a box of size $L$  are considered, where  momenta $p$ of non-interacting single hadrons are multiples of $2\pi/L$. We focus on the system with total momentum zero  with the advantage that the parity is a good quantum number.  

  The resulting operators can be used to extract the discrete energies of eigenstates. These energies  render the scattering   phase shift via the well-known L\"uscher relation \cite{Luscher:1990ux} which originally considered two spin-less particles.  This has been generalized to the scattering of two particles with  spin by various authors, most generaly by \cite{Briceno:2014oea}.  
  
   Certain aspects of constructing the lattice operators for scattering of particles with  spin have already been considered \cite{Berkowitz:2015eaa,Wallace:2015pxa,Gockeler:2012yj,Thomas:2011rh,Dudek:2012gj,Wallace:2015pxa, Moore:2006ng,Briceno:2014oea} before \cite{Prelovsek:2016iyo}.  Despite all previous work, various aspects and proofs were lacking to build a reliable operator   related  to the desired continuum quantum numbers, for example  partial-wave $L$ or single-hadron helicities $\lambda_{1,2}$.

 \section{Single-hadron operators and their transformations}\label{sec:single_hadron}
 
 The single-hadron annihilation operators $H(p)$ need to have the following transformation properties under rotations $R$ and inversion $I$ in order to build two-hadron operators   $H^{(1)}(p)H^{(2)}(-p)$ with desired transformation properties 
  \be \label{2}
R\, H_{m_s}(p)R^{-1}= \sum_{m_s'} D_{m_sm_s'}^s(R^{-1}) H_{m_s'} (R\, p)\;,\qquad  I \, H_{m_s}(p)I = (-1)^P H_{m_s}(-p)\;.~
  \ee
  For a particle at rest $m_s$ is a good quantum number of the spin-component $S_z$.  
 The $m_s$ is generally not a good quantum number for $H_{m_s}(p\not = 0)$; in this case it denotes the  eigenvalue of $S_z$ for the corresponding field $H_{m_s}(0)$, which has good $m_s$. 
 Our two-hadron fields are built from simple (non-canonical)  fields that satisfy  (\ref{2}), for example
  \begin{align}
    \label{6}
   & P(p)=\sum_x \bar q(x) \gamma_5 q(x) e^{ipx}\\
  &  V_{m_s=\pm 1}(p)=\frac{\mp V_x(p)+iV_y(p)}{\sqrt{2}},\  \ V_{m_s=0}(p)=V_z(p),\quad V_i(p)=\sum_x \bar q(x) \gamma_i q(x) e^{ipx}, \ i=x,y,z\nonumber\\
  & N_{m_s=1/2}(p)= {\cal N}_{\mu=1}(p)\;, \ {\cal N}_{m_s=-1/2}(p)= {\cal N}_{\mu=2}(p)\;,   \ \  {\cal N}_\mu(p)\!=\!\sum_x \epsilon_{abc} [q^{aT}(x) C \gamma_5 q^b (x)] ~q^c_{\mu}(x)~e^{ipx}\nonumber
    \end{align} 
        where ${\cal N}_{1,2}$ are the upper two components of Dirac four-spinor ${\cal N}_{\mu=1,..,4}$ in the Dirac basis.    
          
  Another choice of building blocks could be the  canonical hadron fields $H_{m_s}^{(c)}(p)\equiv L(p) H_{m_s}(0)$ obtained after a boost 
   $L(p)$ from $0$ to $p$. Those are less practical since they depend on the hadron  mass $m$, energy $E$ and velocity $v$, for example  $N_{1/2}^{(c)}(p_x)\propto {\cal N}_1(0)+\tfrac{p_x}{m+E}{\cal N}_4$ and $  V_{m_s=1}^{(c)}(p_x)=[-\gamma V_x(p_x)+iV_y(p_x)]/\sqrt{2}$ with $\gamma=(1-v^2)^{1/2}$  \cite{Prelovsek:2016iyo}.  
      
 \section{Transformation properties of two-hadron operators}\label{sec:two_hadron}
 
 The two-hadron operators with zero momentum have good parity $P$ and have to transform as 
 \be
 \label{4}
   R\, O^{J,m_J}(P_{tot}\!=\!0)R^{-1}= \sum_{m_J'} D_{m_Jm_J'}^J(R^{-1}) O^{J,m_J'} (0)\,,\ \   R\in O^{(2)}, \;\;  I O^{J,m_J}(0)I = (-1)^P O^{J,m_J}(0)~.
  \ee
  Such continuum-like operators will present an intermediate step below and the only relevant 
  rotations will be those of the discrete group. 
  
 The continuum rotation group is reduced  to the cubic group on a cubic lattice. It has 24 elements $R\in O$  for integer $J$ and  
 48 elements $R\in O^2$ for half-integer $J$. The number of symmetry elements gets doubled to $\tilde R=\{R,IR\}\in O^{(2)}_h$ when  inversion is a group element. 
 The representations (\ref{4}) with given $J$ and $m_J$ are reducible under  $O^{(2)}_{h}$, so we seek the 
       annihilation operators that transform according to the corresponding irreducible representation (irrep)  $\Gamma$ and row $r$
  \be
  \label{5} 
 R\, O_{\Gamma,r}R^{-1} = \sum_{r'} T^{\Gamma}_{r,r'}(R^{-1})O_{\Gamma,r'} \quad  R\in O^{(2)}  ,\qquad IO_{\Gamma,r}I =(-1)^P O_{\Gamma,r}\;.
     \ee
     The systems with integer $J$ transform according to irreps $\Gamma=A_{1,2}^\pm,~E^\pm,~T_{1,2}^\pm$, while systems with half-integer $J$ according to   $\Gamma=G_{1,2}^\pm~,H^\pm$.  We employ the same conventions for rows in all irreps as in \cite{Bernard:2008ax}, where  explicit representations $T^{\Gamma}_{r,r'}(R)$   are given.

  \section{Two-hadron operators in three methods}\label{sec:three_methods}
  
Here we present  two-hadron operators that transform according to (\ref{5})  derived with three methods.  The continuum-like operators (\ref{4}) will appear as an intermediate  step in two of the methods.  Their  correct transformation properties (\ref{4},\ref{5}) are proven in Appendix of \cite{Prelovsek:2016iyo}. 
      
  \subsection{Projection method}
  
  A projector  to the desired irrep $\Gamma$ and row $r$ on an arbitrary operator   renders  \cite{Dawber:symmetries}
  \be
  \label{O_P} 
    O_{|p|,\Gamma,r,n}=\sum_{\tilde R\in O^{(2)}_h} T^\Gamma _{r,r}(\tilde R)~\tilde RH^{(1),a}(p)H^{(2),a}(-p)\tilde R^{-1}~,\qquad   n=1,..,n_{max}\;.
  \ee 
  The $\tilde R\in O^{(2)}_h $ in (\ref{O_P}) is the operator symbolizing rotations $R$ possibly combined with inversions ($R$ is reserved  for rotations only)    where the action of rotation $R$ or inversion $I$ on all $H=P,V,N$ is given in (\ref{2}).
   The representation matrices  $T^{\Gamma}(\tilde R)$ for all elements $\tilde R$ are listed  for all irreps in  Appendix A of \cite{Bernard:2008ax}.  
   
 The $H^{a}$  are  arbitrary single hadron lattice operators of desired $|p|$ and any  $p$ and $m_s$, for example operators (\ref{6}) or their linear combinations.  We have taken $H^a$ with  all possible combinations of  direction $p$ and polarizations of both particles $m_{s1}$ and $m_{s2}$ (for vectors we chose  $V_x$, $V_y$ and $V_z$ as $H^a$).  For fixed $|p|$, $\Gamma$ and $r$ one can get one or more linearly independent operators  $O_{|p|,\Gamma,r,n}$ which are indicated indexed by $n$. 
 As an illustration we present    resulting $PV$ operators    for irrep $T_1^+$ and $|p|=1$,
  $$O_{|p|=1,T_1^+,r=3,n=1}\propto \!\!\!\sum_{p=\pm e_z}\!\! \text{P}(p) V_z(-p) \;,\ \      O_{|p|=1,T_1^+,r=3,n=2}\propto \!\!\!\!\!\!\sum_{p=\pm e_x,\pm e_y} \!\!\!\!\!\!\! \text{P}(p) V_z(-p)\;,\quad n_{max}=2\;,$$
while others  are listed in   \cite{Prelovsek:2016iyo}.

The projection method  is very general, but it does not offer physics intuition what 
$O_{|p|,\Gamma,r,n}$ with different $n$ represent in terms of the continuum quantum numbers. This will be remedied with the next two methods, that indicate which linear combinations of $O_{n}$ correspond to certain partial waves or helicity quantum numbers.

  \subsection{Helicity method}

The helicity $\lambda$ is the eigenvalue of the helicity operator $h\equiv S\cdotp p/|p|$. It  is a good quantum number for a moving particle with any $p$ (while $m_s$ is a good quantum for $p=0$ and $p\propto e_z$, but not in general).   To obtain a helicity single-hadron  annihilation operator $H_{\lambda}^{h}(p)$, one starts from an operator $H_{m_s=\lambda}(p_z)$ with momentum  $p_z\propto e_z$ that has good $m_s$. This state is   rotated from $p_z$ to desired direction of  $p$ with $R_0^p$ \cite{Jacob:1959at,Thomas:2011rh}
 \be
 \label{13}
 \qquad H_{\lambda}^{h}(p)\equiv  R_0^p ~H_{m_s=\lambda}(p_z) ~(R_0^p)^{-1},\qquad p_z\propto e_z,\quad |p_z|=p~.
\ee  
 The upper index $h$ indicates that  the polarization index  ($\lambda$) stands for the helicity of the particle (not $m_s)$.  
 Simple examples of $H_{p_z,m_s=\lambda}$ can be read-off from (\ref{6}) and the action of $R_0^p$ on them is given by (\ref{2}). 
The arbitrary rotation $R$ rotates the $p$ and  $S$ in the same way,  so the helicity $RH_{\lambda}^{h}(p) R^{-1} \propto H_{\lambda}^{h}(Rp)$
 does not change  \cite{Jacob:1959at,Thomas:2011rh}.
     
 The two-hadron helicity operator   is built from 
 single-hadrons  of given helicities $\lambda_{1,2}$ and arbitrary back-to-back momenta $p$ within a given $|p|$
 \begin{eqnarray}
\label{O_helicity_noP} 
O^{|p|,J,m_J,\lambda_1,\lambda_2,\lambda}&=& \sum_{R\in O^{(2)}} D^J_{m_J,\lambda}(R) ~RH_{\lambda_1}^{(1),h}(p) H_{\lambda_2}^{(2),h}(-p)R^{-1}~.
\end{eqnarray}
The correct transformation properties  (\ref{4}) of the above operator are proven in Appendix of  \cite{Prelovsek:2016iyo}. 
The final   operator with desired parity $P=\pm 1$ is obtained   by parity projection $\tfrac{1}{2}({\cal O}+P I {\cal O} I)$    
\begin{align}
\label{O_helicity} 
O^{|p|,J,m_J,P,\lambda_1,\lambda_2,\lambda}= \frac{1}{2}\!\!\sum_{R\in O^{(2)}} D^J_{m_J,\lambda}(R) ~&RR_0^p~
[H^{(1)}_{m_{s_1}=\lambda_1}(p_z) H^{(2)}_{m_{s_2}=-\lambda_2}(-p_z)\\
&\ +PIH^{(1)}_{m_{s_1}=\lambda_1}(p_z) H^{(2)}_{m_{s_2}=-\lambda_2}(-p_z)I]~(R_0^p)^{-1}R^{-1}~,\nonumber
\end{align} 
where we have expressed $H^{h}$ (\ref{13}) with fields $H_{m_s}(p_z)$ (\ref{6}) that have good quantum number  $m_s$. The actions of inversion  $I$ and the rotation $R$ on the fields $H_{m_s}$ are given in (\ref{2}). One chooses particular $p$ for a fixed $|p|$ and performs rotation $R_0^p$ from $p_z$ to  $p$. There are several possible choices of $R_0^p$, but they lead only to different overall phases for the whole operator (\ref{O_helicity})  \cite{Prelovsek:2016iyo}, which is irrelevant. 

As an illustration we present  $PV$ operators in the $J^P=1^+$ channel 
$$O^{|p|=1,J=1,m_J=0,P=+,\lambda_V=0}\propto \!\!\!\sum_{p=\pm e_z} \text{P}(p) V_z(-p) ~,\ \       O^{|p|=1,J=1,m_J=0,P=+,\lambda_V=1}\propto \!\!\!\!\sum_{p=\pm e_x,\pm e_y} \!\!\!\!\text{P}(p) V_z(-p)~, $$
where the simplest choice $p=p_z=(0,0,1)$ and    $R_0^p=\mathbf{1}$  in (\ref{O_helicity}) can be used. 
 
 The helicity operators (\ref{O_helicity})   would correspond to irreducible representations only for the continuum rotation group. These represent reducible representation under the  discrete group $O^{(2)}$. In simulations  it is convenient to  employ operators, which transform according to irreducible representations $\Gamma$ and row $r$ of $G=O^{(2)}$. Those are obtained    from   $O^{J,m_J}$ by the subduction  \cite{Dudek:2010wm,Edwards:2011jj} 
  \be
  \label{14}
  O_{|p|,\Gamma,r}^{[J,P,\lambda_1,\lambda_2,\lambda]}=\sum_{m_J}  {\cal S}^{J,m_J}_{\Gamma,r} O^{|p|,J,m_J,P,\lambda_1,\lambda_2,\lambda}\;.
       \ee 
The subduction coefficients  ${\cal S}$   are given in   Appendices  of \cite{Dudek:2010wm,Edwards:2011jj} for all irreps\footnote{ For the $T_{1}$ we choose rows as $x,y,z$ which differs from   \cite{Dudek:2010wm,Edwards:2011jj},   and those  ${\cal S}$ are listed in Appendix  of    \cite{Prelovsek:2016iyo}. }. 
            We expect that the subduced operators  $O_{|p|,\Gamma,r}^{[J,P,\lambda_1,\lambda_2,\lambda]}$ will carry the memory of 
      continuum $J,\lambda_1,\lambda_2,\lambda$ and will dominantly couple to eigenstates with these quantum numbers \cite{Dudek:2010wm}.   
 
  \subsection{Partial-wave method}  
 
Often one is interested in the scattering of two hadrons in a given partial wave $L$. 
   The orbital angular momentum $L$ and total spin $S$ are not separately conserved, so several $(L,S)$ combinations can render the same $J$, $m_J$ and $P$, which are good quantum numbers.   Nevertheless, the $L$ and $S$ are valuable physics quantities to label continuum annihilation field operator:  
   \be
\label{O_LS}
O^{|p|,J,m_J,S,L}=\sum_{m_L,m_S,m_{s1},m_{s2}} C^{Jm_J}_{Lm_L,Sm_S}C^{Sm_S}_{s_1m_{s1},s_2m_{s2}} \sum_{R\in O} Y^*_{Lm_L}(\widehat{Rp}) H^{(1)}_{m_{s1}}(Rp)H^{(2)}_{m_{s2}}(-Rp)~.
\ee
  The operator has parity $P=P_1P_2(-1)^L$ and its correct transformation property under rotation (\ref{4}) is demonstrated in Appendix of \cite{Prelovsek:2016iyo}.   The operator was considered for nucleon-nucleon scattering in \cite{Berkowitz:2015eaa}  (this reference uses $Y_{Lm_L}$ where we have $Y_{Lm_L}^*$).  The $C$ are Clebsch-Gordan coefficients, $p$ is an arbitrary momentum with desired $|p|$, $Y_{Lm_L} $ is a spherical harmonic. Simple choices of one-particle   operators $H$ are listed in (\ref{6}).      Here is an example of $PV$ operators with the same $J$ and $S$ 
  \begin{align}
  O^{|p|=1,J=1,m_J=0,L=0,S=1}&\propto \!\!\!\!\!\!\sum_{p=\pm  e_x,\pm e_y, \pm e_z}\!\!\!\!\!\!\! \text{P}(p) V_z(-p) \;,\nonumber\\
     O^{|p|=1,J=1,m_J=0, L=2,S=1}&\propto \!\!\! \sum_{p=\pm  e_x,\pm e_y } \!\!\!\!\!\!\text{P}(p) V_z(-p) - 2 \!\! \sum_{p=\pm e_z } \!\text{P}(p) V_z(-p)\;. \nonumber
  \end{align}
          
 The  operators that transform according to irrep $\Gamma$ and row $r$ of $G=O^{(2)}$ are obtained from $O^{J,m_J}$ by the subduction  \cite{Dudek:2010wm} using   the same coefficients ${\cal S}$ as in (\ref{14})
  \be
  \label{15}
  O_{|p|,\Gamma,r}^{[J,S,L]}=\sum_{m_J}  {\cal S}^{J,m_J}_{\Gamma,r} O^{|p|,J,m_J,S,L}\;.
       \ee  
      One expects that the subduced operators  $O_{|p|,\Gamma,r}^{[J,S,L]}$ carry the memory of 
      continuum $J,S,L$ and  dominantly couple to eigenstates with these  quantum numbers. 
     
  \section{Operators  for $PV$, $PN$, $VN$ and $NN$ scattering}\label{sec:results}

  The explicit expressions for operators $H^{(1)}(p)H^{(2)}(-p)$ for $PV$, $PN$, $VN$ and $NN$ scattering in three methods are collected 
  in our longer publication \cite{Prelovsek:2016iyo}. Operators for the lowest two momenta  $|p|=0,1$ (in units of $2\pi/L$)  are presented for all irreducible represenations. Expressions for higher  $|p|$ can be obtained using the general expressions (\ref{O_P}), (\ref{O_helicity},\ref{14}) and (\ref{O_LS},\ref{15}) for the three methods, respectively.  
  
  Let us illustrate the results on  operators for $PV$ scattering in the irreducible representation $T_1^+$. This irrep contains states with positive parity and $J=1$ (as well as  $J\geq 3)$. The operators for the  row $r=3$ ($z$-component) are  
 \bea 
 \label{T1p}
 |p|=0:\quad &&O_{T_1^+}=O_{T_1^+}^{[J=1,L=0,S=1 ]}= \text{P}(0) V_z(0)\\
 &&  ~\quad\nonumber \\
|p|=1:\quad&& O_{T_1^+,n=1}=\text{P}(e_z) V_z(-e_z)+\text{P}(-e_z) V_z(e_z)   \nonumber \\
&& O_{T_1^+,n=2}=\text{P}(e_x) V_z(-e_x)+\text{P}(-e_x) V_z(e_x)+\text{P}(e_y) V_z(-e_y)+\text{P}(-e_y) V_z(e_y)  \nonumber \\ 
&& O_{T_1^+}^{[J=1,P=+, \lambda_{V}= \pm 1,\lambda_P=0]}=  O_{T_1^+,n=2}    \nonumber\\
&& O_{T_1^+}^{[J=1,P=+, \lambda_{V}=0 ,\lambda_P=0]}=  O_{T_1^+,n=1}      \nonumber \\
&&O_{T_1^+}^{[J=1,L=0,S=1 ]}= O_{T_1^+,r=1,n=1} + O_{T_1^+,n=2}   \nonumber \\
&& O_{T_1^+}^{[J=1,L=2, S=1] }= -2 ~ O_{T_1^+,r=1,n=1} + O_{T_1^+,n=2}~.
\eea 
The projection method renders two linearly independent interpolators $O_{T_1^+,n=1,2}$  at $|p|=1$ for each row. This method does not tell  which  
partial-waves and single-hadron helicities correspond to each operator. This is remedied by the partial-wave and helicity\footnote{We consider helicity only for the case $|p|\not =0$. } operators.
The expressions $O_{T_1^+}^{[J=1,L,S=1 ]}$ (\ref{T1p}) indicate which linear combinations of $O_{T_1^+,n}$  need to be employed to study $L=0$ or $L=2$ partial waves. Note that both partial waves inevitably contribute to the same $J^P=1^+$ channel even in the continuum $PV$ scattering with $S=1$. 
The $O_{T_1^+}^{[J=1,P=+, \lambda_{V},\lambda_P]}$ (\ref{T1p})  indicate that $O_{T_1^+,n=1}$ is  relevant for $\lambda_V=0$, while  $O_{T_1^+,n=2}$ is relevant for  $|\lambda_V|=1$.   All methods lead to two linearly independent operators that are consistent with each other.  

In general, one or several linearly independent operators $O_{\Gamma,r,n}$  arise from the projection method for $PV$, $PN$, $VN$ and $NN$ scattering in given irrep $\Gamma$.  The explicit partial-wave  operators  in \cite{Prelovsek:2016iyo} indicate which linear combinations of $O_{\Gamma,r,n}$ are relevant to study  the channel $(J,L,S)$. The expressions for helicity operators in \cite{Prelovsek:2016iyo} tell us which linear combinations of $O_{\Gamma,r,n}$ are relevant to study scattering with given $(J,P,\lambda_1,\lambda_2)$. All three methods render the same number of linearly independent operators, which also agrees with the number based on   \cite{Moore:2006ng}. The explicit expressions for operators with $|p|=0,1$ also show that the three methods lead to the consistent results, i.e. operators from partial-wave or projection method can always be expressed as a linear combination of operators from projection method.  
 
\section{Conclusions}

We   construct  two-hadron interpolators which are relevant to simulate  $PV$, $PN$, $VN$ or $NN$ scattering using quantum field theory on the lattice. 
 Here $P$, $V$ and $N$ denote pseudoscalar, vector and nucleon, respectively.    The focus is on the case with total-momentum zero where parity   is a good quantum number.   The projection method is a general mathematical tool which leads to one or several operators $O_{\Gamma,r,n}$  that transform according to given irrep $\Gamma$ and row $r$, but it does not give much insight on the underlying continuum quantum numbers. The partial-wave  and the helicity methods indicate which linear combinations $O_{\Gamma,r,n}$ of various $n$ have to be employed in the in order to enhance couplings to the states with desired continuum quantum numbers.  
The partial-wave method renders operators $O_{\Gamma,r}^{[J,S,L]}$ with enhanced couplings to two-hadron states in partial wave $L$, total spin $S$ and total angular momentum $J$. The helicity method provides operators $O_{\Gamma,r}^{[J,P,\lambda_1,\lambda_2]}$  where each hadron has good helicity $\lambda_{1,2}$.  All three formally independent methods lead  to  consistent results.
     
     \vspace{0.4cm}
     
     {\bf Acknowledgments}
We acknowledge discussions with  R. Brice\~no, J. Dudek, R. Edwards, A. Nicholson, M. Padmanath and A. Walker-Loud. We thank M. Padmanath for insightful discussions on   $PN$ interpolators. This work is supported in part by the  Slovenian Research Agency ARRS and by the Austrian Science Fund FWF:I1313-N27. S.P. acknowledges support from U.S. Department of Energy Contract No. 
DE-AC05-06OR23177, under which Jefferson Science Associates, LLC, manages and operates Jefferson Laboratory.

 \bibliography{bib/Lgt}
\bibliographystyle{h-physrev4}

\end{document}